\documentclass[preprint2]{aastex}
\usepackage{epsfig}
\usepackage{psfig}
\usepackage{lscape}

\newcommand{\psr}{PSR~B2224+65}

\shorttitle{XMM observation of \psr}
\shortauthors{Hui et al.}

\begin{document}

\title{\emph{XMM-Newton} observation of \psr\ and its jet}

\author{C. Y. Hui\altaffilmark{1}, R. H. H. Huang\altaffilmark{2},
L. Trepl\altaffilmark{3}, N. Tetzlaff\altaffilmark{3}, J. Takata\altaffilmark{4}, E. M. H. Wu\altaffilmark{4} and K. S. Cheng\altaffilmark{4}}

\altaffiltext{1}{Department of Astronomy and Space Science, Chungnam
National University, Daejeon 305-764, Korea}
\altaffiltext{2}
{Institute of Astronomy and Department of Physics, National Tsing Hua University, Hsinchu, Taiwan}
\altaffiltext{3}
{Astrophysikalisches Institut und Universit\"ats-Sternwarte, Universit\"at Jena, Schillerg\"a$\beta$chen 2-3, 07745 Jena, Germany}
\altaffiltext{4}
{Department of Physics, University of Hong Kong, Pokfulam Road, Hong
Kong}


\begin{abstract}
We have investigated the pulsar \psr\ 
and its X-ray jet with XMM-Newton. Apart from the long X-ray 
jet which is almost perpendicular to the direction of proper motion, a putative
extended feature at the pulsar position, which oriented in the opposite direction 
of the proper motion, is also suggested by this deep X-ray imaging.
Non-detection of any coherent X-ray pulsation 
disfavors the magnetospheric origin of the X-rays observed from the position
of \psr\ and hence suggest that the interpretation of pulsar wind nebula is more
viable. We have also probed the origin of \psr\ 
and identified a runaway star, which possibly originated from the Cygnus OB9
assoication, as a candidate for the former binary companion of the neutron star's
progenitor.
\end{abstract}


\section{INTRODUCTION}

While the signature of rotation-powered pulsars is
their pulsed emission, bulk of the rotational energy 
is going to power a relativistic outflow of magnetohydrodynamic
wind particles with Lorentz factors up to $\gamma\sim10^8$. Although 
pulsar winds cannot be observed directly, their presence can
be inferred through their interaction with the interstellar surroundings.
If the wind reaches a termination shock, a pulsar wind nebula (PWN) is
formed. The term PWN had been originally referred to the shocked emission 
associated with young and powerful pulsars (e.g. Crab) where the
wind is effectively confined by the magnetic fields of the surrounding supernova remnants  
and relativistic leptons reach high energy densities
which make these nebulae usually luminous synchrotron sources in the
radio and X-ray regimes.

A different class of PWNe which is associated with
older and less powerful pulsars is bow-shock nebulae.
These nebulae are not embedded in supernova remnants but interact
directly with the interstellar medium (ISM). If the pulsar has a
high velocity, the nebula is confined by the ram pressure of the
ISM and forms the typical bow-shock structure. The bow-shock size
is determined by the balance between the wind and the ISM ram pressure.
Even in this old and less powerful pulsars, there is a significant fraction of the whole
pulsar spin-down power which is going to feed the pulsar wind (see Bucciantini \& Bandiera 2001).

Among all the pulsars with PWNe detected, \psr\ is one of the most interesting system. 
This pulsar has a rotational period and a period derivative of 
$P\sim0.68$~s and $\dot{P}\sim9.6\times10^{-15}$~s~s$^{-1}$, respectively. 
This suggests \psr\ has a spin-down age of $1.13\times10^{6}$ years. The spin parameters 
imply a spin-down luminosity and a surface dipole magnetic field strength of 
$\dot{E}\sim1.2\times10^{33}$~erg~s$^{-1}$ 
and $B\sim2.6\times10^{12}$~G respectively. 
Concerning its distance, while the radio dispersion measure implies a 
value of 1.86 kpc (cf. Cordes \& Lazio 2002), 
Chatterjee \& Cordes (2004) found that by modeling the head of the shocked 
emission in H$\alpha$ the pulsar distance could be as close as 1 kpc.

Hui \& Becker (2007) 
have examined its X-ray properties with the Chandra data taken in October 2000. 
Brief results of the same observation have also been reported by Wong et al. (2003) and 
Zavlin \& Pavlov (2004). Despite the photon 
statistics is limited in this data set, the spectral analysis suggests that the bulk of the X-ray 
emission from \psr\ is non-thermal. An independent analysis that includes also the second epoch Chandra
observations taken in August-October 2006 has confirmed the non-thermal nature of the X-rays detected at 
the pulsar position (Johnson \& Wang 2010). 
Nevertheless, the large frame time (i.e. 3.2 sec) of these Chandra observations do 
not allow us to constrain the temporal properties of the pulsar. 
Therefore, with the existing data, we cannot determine whether the 
bulk of these observed X-rays are originated from the pulsar magnetosphere or from the PWN.  

The proper motion of \psr\ is among the largest in the currently known pulsar population with 
a transverse velocity of $v=865(d/1~{\rm kpc})$~km~s$^{-1}$ (Harrison, Lyne \& Anderson 1993). 
Its shock interaction with the ISM produces the most well-known 
\emph{Guitar Nebula} as observed in H$\alpha$ (Cordes et al. 1993). Although no evidence for the 
X-ray emission can be found in the region of the H$\alpha$ nebula, 
an X-ray jet apparently associated with \psr\ has been detected by Chandra 
(Hui \& Becker 2007; Wong et al. 2003). The most peculiar 
fact of this jet is that its orientation deviates by $\sim118^{\circ}$ from the pulsar's proper 
motion direction. Its X-ray spectrum is also found to be non-thermal. 

Kargaltsev \& Pavlov (2008) have argued that this jet-like feature may be associated with a nearby 
point source instead of with \psr. However, this possibility has been ruled out by a recent analysis of 
the multi-epoch Chandra observations. Johnson \& Wang (2010) have found that the sharp leading edge have 
a proper motion consistent with that of the pulsar, which secures the association between the jet and \psr. 
Different pictures have been proposed to explain this X-ray jet (e.g. Bandiera 2008; 
Johnson \& Wang 2010). However, there is still no consensus on its nature. 


To further investigate the nature of the X-rays from the position of \psr\ and 
its enigmatic jet, a dedicated follow-up observation with improved temporal resolution and photon statistic is needed. 
In this paper, 
we present a deep XMM-Newton observation in order to study this extreme system in details. In 
\S2, we describe the details of the XMM-Newton observation and the data reduction. In \S3 and \S4, we present the results 
of the X-ray analysis and the search for the $\gamma-$ray emission from this system respectively. 
Finally, we discuss the physical implications of the observed results and search for the birth place of this fast-moving 
neutron star in \S5.  

\section{X-RAY OBSERVATION WITH XMM-NEWTON}

 \psr\ was observed by XMM-Newton on 13-14 July 2009 (Observation ID: 0604420101) with an 
 exposure time of 75~ks. The mean epoch of this observation is at MJD 55025. 
 The data used in this investigation were obtained with the {\bf E}uropean {\bf P}hoton
 {\bf I}maging {\bf C}amera (EPIC) aboard XMM-Newton (Jansen et al.~2001). EPIC consists of
 two Metal Oxide Semiconductor (MOS1/2) CCD detectors (Turner et al.~2001) of which half of
 the beam from two of the three X-ray telescopes is reflected to. The other two halves of the
 incoming photon beams are reflected to a grating spectrometer (RGS) (den Herder et al.~2001).
 The third of the three X-ray telescopes is dedicated to expose the PN CCD detector solely
 (Str\"uder et al.~2001). 
 The PN CCD was operated in small-window mode with a thin filter to block optical stray light.
 All the events recorded by PN camera are time 
 tagged with a temporal resolution of 5.7 ms, which enable us to constrain the temporal properties of this pulsar 
 in X-ray regime for the first time. On the other hand, the MOS1/2 CCDs were setup to operate in 
 full-window mode with a thin filter in each camera. 

 The aimpoint of the satellite in this observation is RA=$22^{\rm h}25^{\rm m}52.43^{\rm s}$
 and Dec=$+65^\circ 35' 34.0''$ (J2000). With the most updated instrumental calibration, we 
 generate the event lists from the raw data obtained from all EPIC instruments 
 with the tasks \emph{emproc} and \emph{epproc} version 9.0.0 of the XMM {\bf S}cience {\bf A}nalysis {\bf S}oftware 
 (XMMSAS version 9.0.0). Examining the raw data from the PN CCD, we did not find any 
 timing anomaly which was observed in many of the XMM-Newton data sets 
 (cf. Hui \& Becker 2006 and references therein). This provides us with opportunities for 
 an accurate timing analysis. We then created  filtered event files for the energy range 0.5 keV 
 to 10 keV for all EPIC instruments and selected only those events for which the pattern was 
 between $0-12$ for MOS cameras and $0-4$ for the PN
 camera. We further cleaned the data by accepting only the good times when sky background was low and
 removed all events potentially contaminated by bad pixels. After the filtering, the effective 
 exposures are found to be 46~ks for each MOS camera and 32~ks for PN. 
 
\section{ANALYSIS OF XMM-NEWTON OBSERVATION}

\subsection{Spatial Analysis}

\begin{figure*}
\centerline{\psfig{figure=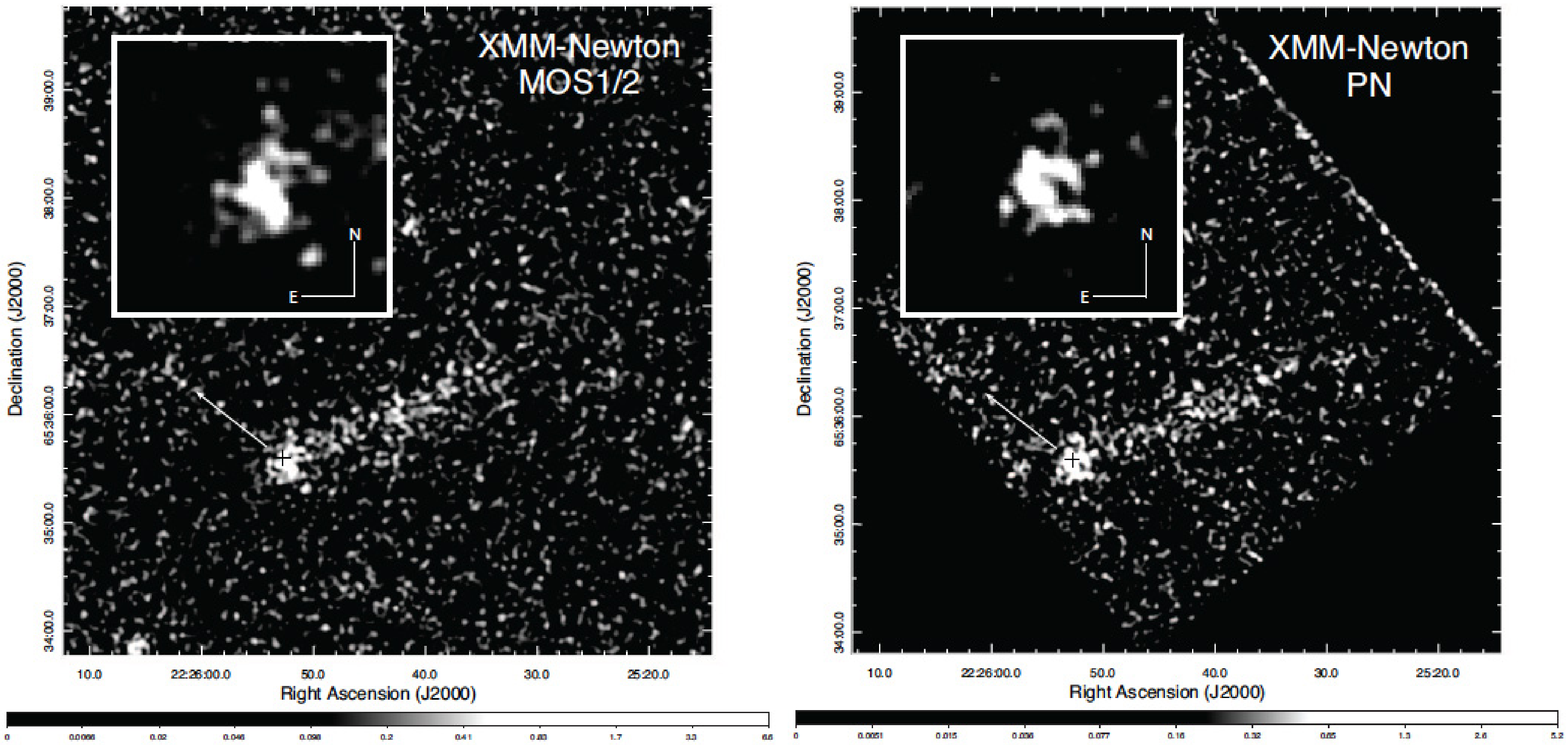,width=18cm,clip=}}
\caption[]{The $6'\times6'$ field-of-view of \psr\ and its jet as observed by XMM-Newton MOS1/2 cameras ({\it left panel}) and 
PN camera ({\it right panel}) in the energy band of $0.5-10$~keV. The proper motion direction and the proper motion corrected 
position of the pulsar is illustrated by the black cross and white arrow respectively. The insets show the $1'\times1'$ 
close-up view around \psr. All images have been smooth by a Gaussian with $\sigma=3''$.} 
\label{xmm_img}
\end{figure*}

X-ray images of the $6'\times6'$ field around \psr\ as obtained 
by MOS1/2 and PN camera are displayed in Figure~\ref{xmm_img}. 
Both image have been firstly produced with a binsize of $1''$ which
is subsequently smoothed with a Gaussian with a kernel size 
of $\sigma=3''$. Both the pulsar and the jet can be clearly seen. 

\begin{figure*}
\centerline{\psfig{figure=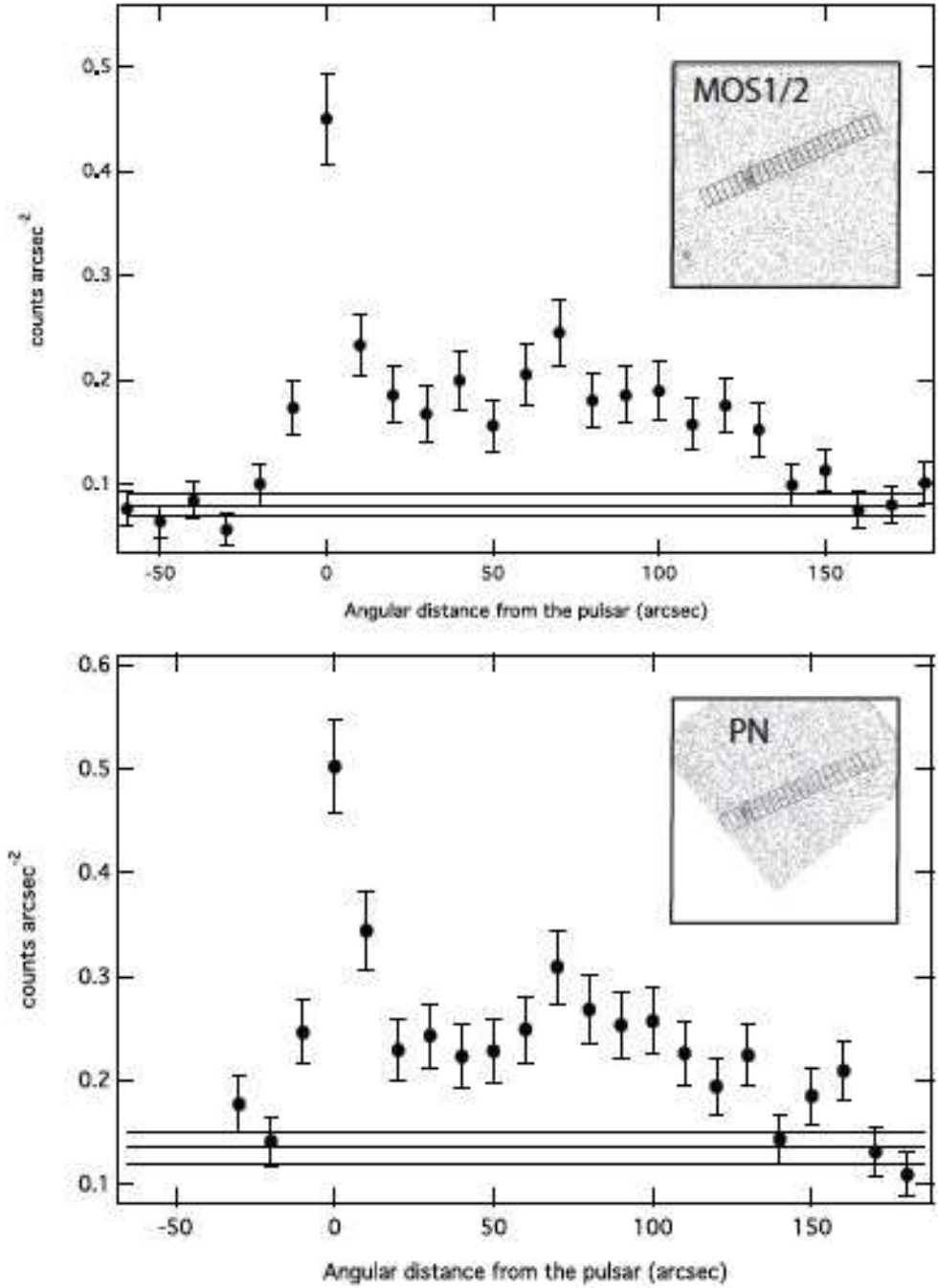,width=14cm,clip=}}
\caption[]{The X-ray brightness profile in the energy band of $0.5-10$~keV along the orientation of the jet as observed 
by XMM-Newton MOS1/2 cameras ({\it upper panel}) and PN camera ({\it lower panel}). The insets show the bins used in 
computing the profiles in the corresponding data sets. Each bin has a size of $25''\times10''$. The average 
background level and its $1\sigma$ deviation are indicated by horizontal lines which were calculated by sampling 
the source-free regions in the $6'\times6'$ field-of-view of the corresponding data sets as shown in Figure~\ref{xmm_img}.}
\label{brightness}
\end{figure*}

For investigating the spatial variation 
of the X-ray intensity of the jet, we computed the brightness profiles along its orientation with 
MOS1/2 and PN data which are shown in Figure~\ref{brightness}. We 
estimate the counts in the consecutive boxes with a size of $25''\times10''$ from the raw images 
with a binsize of $1^{"}$ along the jet. The orientation of these sampling regions are shown in 
the insets of Figure~\ref{brightness}. To estimate the background level, we have sampled the source-free 
regions around \psr\ within a $6'\times6'$ field-of-view. The average background level and its 
$1\sigma$ uncertainties are indicated by the horizontal solid line and dotted lines in Figure~\ref{brightness}
respectively. Apparently, the jet extends for $\sim2.5'$ before it falls 
to the estimated background level. Apart from the peak corresponds to \psr, another peak at 
$\sim70''$ from the pulsar position is also observed in both MOS1/2 and PN brightness profiles. 
Hereafter, we refer this as a ``knot-like" feature. 
To determine the position of the knot-like feature on the jet, we have performed source detection individually on each 
EPIC data set with the aid of the XMMSAS task \emph{edetect\_chain}. The mean X-ray position inferred from this 
observation is RA=$22^{\rm h}25^{\rm m}42.24^{\rm s}$ Dec=$+65^{\circ}36^{'}00.0^{''}$ (J2000) 
with a resultant uncertainty of $5^{''}$ by 
combining the statistical errors inferred from each camera in quadrature.

To further constrain the nature of the knot-like feature, we make use of the archival data obtained 
by the Hubble Space Telescope (HST). Utilizing the DAOphot catalog (version 4.0) released 
on the \emph{Hubble Legacy Archive}\footnote{http://hla.stsci.edu/hlaview.html}, we have identified 
a possible optical counterpart with a magnitude of $R\sim22.31$ at 
RA=$22^{\rm h}25^{\rm m}42.14^{\rm s}$ Dec=$+65^{\circ}36^{'}01.2^{''}$ (J2000) which locates within  
the X-ray positional uncertainty of the knot\footnote{The position of this optical counterpart is inferred 
by a HST observation performed on 2001-11-22 (Proposal ID: 9129)}. 
The root mean square (rms) positional uncertainty is $\sim0.2^{''}$ of the optical source 
in each coordinate based on the
astrometric correction with 53 reference sources in Guide Star Catalog 2 (Lasker et al. 2008). 
Assuming the optical source is 
associated with the X-ray knot, we can probe its possible nature together with X-ray spectral analysis 
(see \S3.2). 

With the archival data obtained by Chandra X-ray Observatory, it is 
possible for us to place a constraint on the knot's proper motion. We utilized the
{\bf C}handra {\bf I}nteractive {\bf A}nalysis of {\bf O}bservations software (CIAO~4.3) for the re-analysis.
We have reprocessed all the available data of \psr\ (Obs. IDs 755; 6691; 7400) obtained by the 
Advanced CCD Imaging Spectrometer (ACIS) onboard Chandra with the most updated calibration CALDB (ver.~4.4.5).
Different from the analysis reported by Johnson \& Wang (2010) and
Hui \& Becker (2007), we have also applied the subpixel event repositioning during reprocessing the data in order
to improve the positional accuracy of each event (cf. Li et al. 2004). For this analysis, we considered only the 
Obs. IDs 755 and 7400 which have an epoch separation of $\sim6$~yrs. The knot can be detected in both images 
with a wavelet source detection algorithm. Since Johnson \& Wang (2010) found that the rotation between these images 
is negligible, simple translational astrometric corrections have been applied. The proper motion of the knot 
inferred from these two frames is $\mu_{\rm RA}=-1.0\pm0.4$~arcsec/yrs and $\mu_{\rm RA}=0.18\pm0.15$~arcsec/yrs. 
These errors ($1\sigma$) are computed by combining 
the X-ray positional errors of the knot in both frames and the corresponding astrometric uncertainties in 
quadrature. The error budget is dominated by the uncertainties in determining the positions of the knot. 
In view of the large errors, the proper motion of the knot remains unconstrained.

\begin{figure*}
\centerline{\psfig{figure=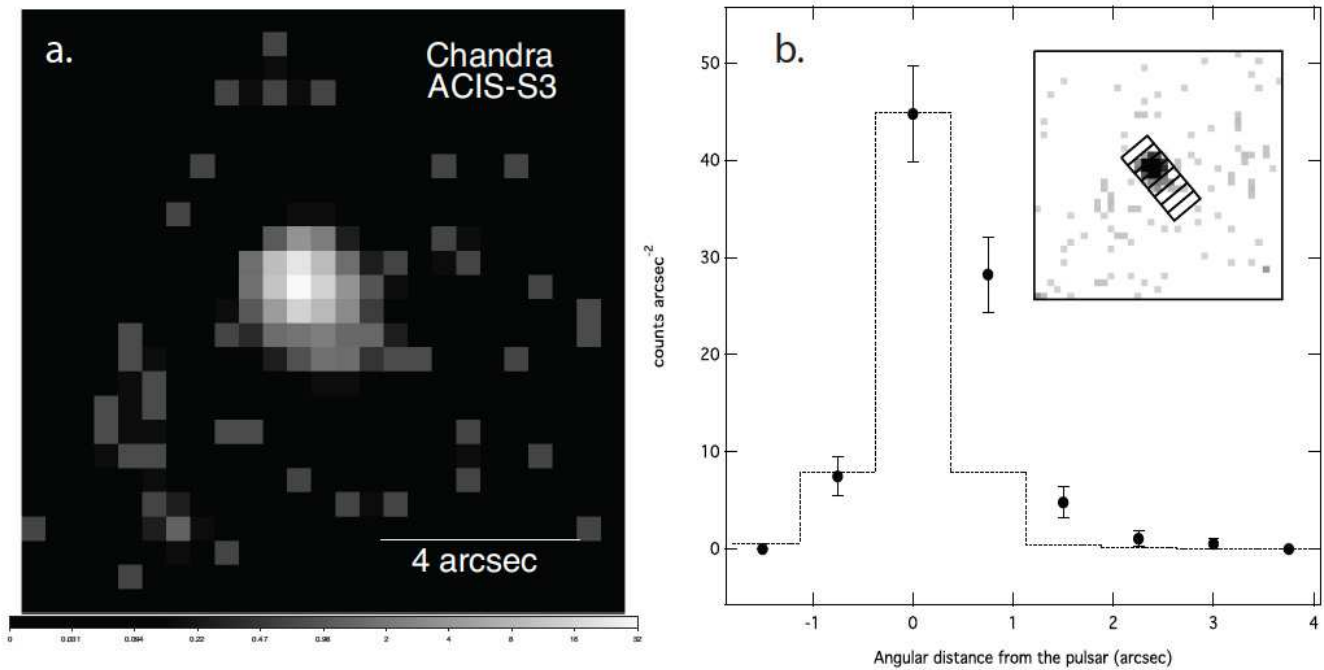,width=18cm,clip=}}
\caption[]{{\bf a.} The X-ray image in the energy band $0.5-8$~keV centered on \psr\ 
by merging the Chandra ACIS data obtained in different epoch. Top is north and left is east. 
{\bf b.} X-ray brightness profile in $0.5-8$~keV as sampled from eight bins along the 
orientation as illustrated in the inset. 
The dash line indicates the expected profile for a point source.}
\label{cxc}
\end{figure*}

The insets of Figure~\ref{xmm_img} show the $1'\times1'$ close-up view of \psr\ in the corresponding cameras. 
The X-ray emission from the pulsar position appears to be slightly elongated for $\sim8''$ behind
proper motion direction as illustrated by the arrow in Figure~\ref{xmm_img}.  In both MOS1/2 and PN images, 
this putative feature apparently has the same orientation. However, the
elongation of this feature is close to the edge of the spatial resolution of XMM-Newton. This has motivated 
us to reexamine the archival Chandra data which can offer an improved angular resolution. 
We produced a merged 
epoch X-ray image after correcting the relative astrometric offsets among each data set. The epoch-merged 
raw X-ray image of \psr\ is shown in Figure~\ref{cxc}a. Figure~\ref{cxc}b shows the X-ray brightness profile 
as measured by eight consecutive boxes with the orientation illustrated in the inset. 
Interestingly, both image and the brightness profile suggest a slight deviation from a
point source with deformation toward a similar direction as observed by XMM-Newton. The extent of 
$\sim3''$ is observed in the ACIS data, which can possibly be the part of the highest surface brightness, though it is 
not conclusive with the current data (see \S5 for a further discussion). 

\subsection{Spectral Analysis}

With the aid of the XMMSAS task {\it epatplot}, 
all the EPIC data are found to be not affected by CCD pileup.
For \psr, we extracted its source
spectrum from circles with a radius of $20''$ centered at the point source position in both MOS1/2 and PN cameras
respectively. This choice of extraction regions corresponds to the encircled energy fraction of $\sim75\%$ and $\sim80\%$
in MOS1/2 and PN respectively
\footnote{http://xmm.esac.esa.int/external/xmm\_ user\_ support/documentation/uhb/node17.html}. 
For the jet, we sampled its source spectrum from a $122''\times36''$ 
box centered at RA=$22^{\rm h}25^{\rm m}40.69^{\rm s}$ Dec=$+65^{\circ}36^{'}07.64^{''}$ (J2000) and 
the longer side oriented $66^{\circ}$ east from the north so as to cover the whole jet. 
The background spectra were extracted from the nearby regions which are source-free and with sufficient size to
enable a less biased sampling. Response files were computed for all datasets by using the XMMSAS tasks {\it rmfgen} and 
{\it arfgen}. After the background subtraction, we have $74\pm11$~cts (MOS1), $67\pm10$~cts (MOS2), $189\pm17$~cts (PN) 
collected from \psr\ and $183\pm18$~cts (MOS1), $212\pm18$~cts (MOS2), $453\pm30$~cts (PN) collected from the jet respectively.

We have firstly examined the data obtained from each camera individually and found that 
all the spectral parameters inferred from different cameras are consistent within $1\sigma$ uncertainties. In order to tightly  
constrain the spectral parameters, we fitted the data obtained from three cameras simultaneously for 
the individual analysis of \psr\ and its jet. According to the photon statistics, we grouped each spectrum dynamically so 
as to achieve the same signal-to-noise rate in each dataset. The adopted spectral binning ensures a 
sufficiently small errors in each energy bin for better discriminating the competing scenarios. 
All the spectral fits were performed with 
XSPEC 12.6.0 in the energy range of $0.5-10$~keV. All the quoted errors are $1\sigma$ for 1 parameter of 
interest (i.e. $\Delta\chi^{2}=1.0$ above the minimum). 

\begin{figure*}
\centerline{\psfig{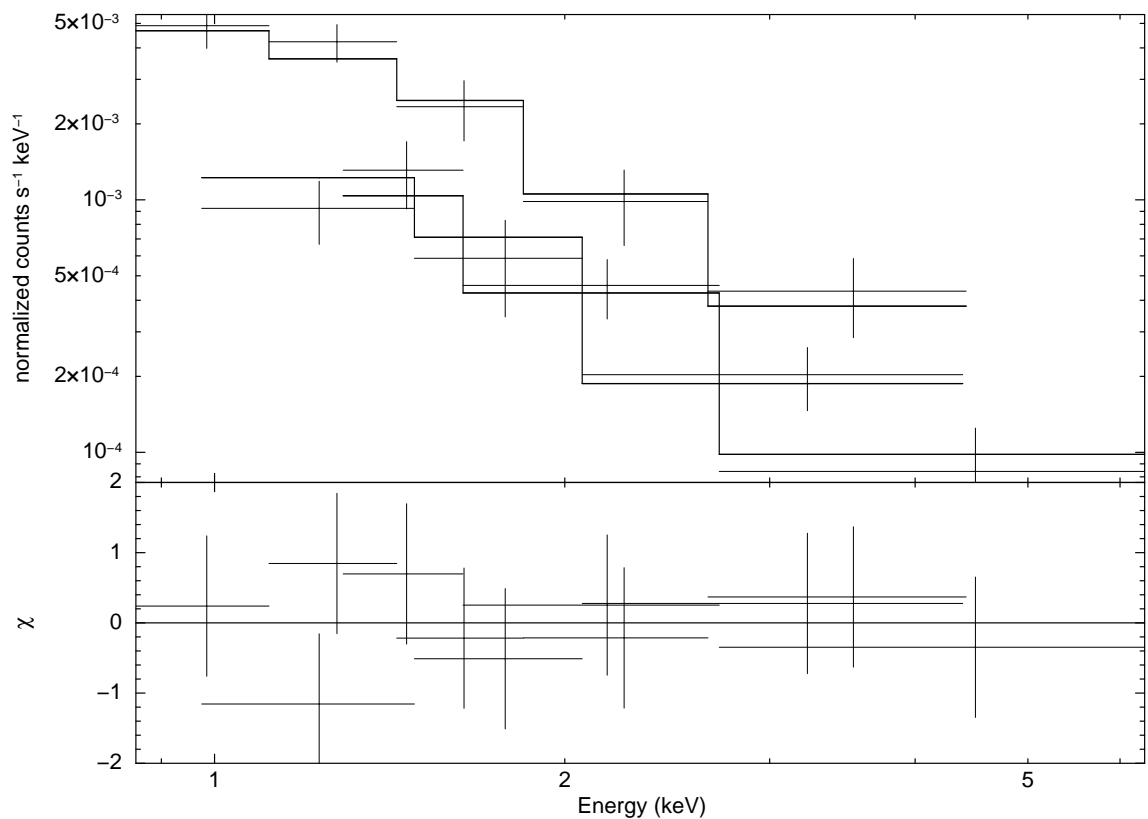}}
\caption[]{The energy spectrum of X-ray emission from the position of \psr\ as observed 
with the PN (upper spectrum) and MOS1/2 detectors (lower spectra) and 
simultaneously fitted to an absorbed power-law model ({\it upper panel}) and contributions to the $\chi^{2}$ fit statistic 
({\it lower panel}).}
\label{psr_spec}
\end{figure*}

For \psr, we found that a simple absorbed power-law model yields a goodness-of-fit of 
$\chi^{2}=3.35$ with 8 d.o.f.. Examining the power-law with different spectral binning factor, the model 
remains to provide a very good description of the data. The best-fit power-law model and the residuals for the pulsar spectrum are 
shown in Figure~\ref{psr_spec}. This model yields a column density of $n_{H}=(1.3^{+2.2}_{-1.2})\times10^{21}$~cm$^{-2}$, a 
photon index of $\Gamma_{X}=2.0^{+0.5}_{-0.4}$ and a normalization at 1 keV of 
$(6.7^{+3.8}_{-2.0})\times10^{-6}$~photons~keV$^{-1}$~cm$^{-2}$~s$^{-1}$. 
We have also examined whether the X-rays from \psr\ can be described in a thermal-dominant emission scenario by 
fitting the spectrum with a blackbody model. The best-fit model yields a column density of $n_{H}<4.6\times10^{20}$~cm$^{-2}$, 
a temperature of $kT=0.5\pm0.1$~keV and an emitting radius of $R_{\rm bb}=(14.1^{+2.8}_{-2.3})\times d_{\rm 1 kpc}$~m, 
where $d_{\rm 1 kpc}$ 
at the unit of 1~kpc. The inferred emitting region is smaller than the radius of the polar cap (i.e. $R_{\rm pc}=175$~m) as 
defined by the last open field lines of a dipolar magnetic field. Also, the goodness-of-fit resulted from the best-fit 
blackbody model is somewhat worse ($\chi^{2}=9.28$ with 8 d.o.f.) in comparison with the power-law fit. 
Examining the fitting residuals, some systematic deviations are found between the observed spectrum and the best-fit blackbody 
model. In view of these, the blackbody model is not favored. Therefore, we conclude that the simple absorbed power-law model
is favored in modeling the observed spectrum of \psr, which agrees with the inference drawn by Hui \& Becker (2007)
and Johnson \& Wang (2010).

For the jet, we found that a power-law model yields a column density of $n_{H}=(2.7^{+1.2}_{-0.8})\times10^{21}$~cm$^{-2}$, a
photon index of $\Gamma_{X}=1.2\pm0.1$ and a normalization at 1 keV of
$(9.1^{+2.3}_{-1.7})\times10^{-6}$~photons~keV$^{-1}$~cm$^{-2}$~s$^{-1}$. The model results in a goodness-of-fit of $\chi^{2}=38.37$ 
with 40 d.o.f. The inferred column density from fitting the jet spectrum is consistent with that inferred from 
the pulsar spectrum. On the other hand, in examining whether a thermal 
scenario is capable to describe the observed data, we attempted to 
fit a thermal bremsstrahlung model to the spectrum of the jet. However, we found that it results in an unphysically high temperature 
of $kT\sim180$~keV. 
Therefore, our investigation further supports a non-thermal origin of the X-rays from the jet. 

In order to constrain the spectral properties of this system more tightly, 
we follow the method adopted by Johnson \& Wang (2010) to jointly fit 
individual power-law models for the X-ray spectra of the pulsar and 
the jet with the column density in the individual model tied together. 
The results are summarized in Table~\ref{spec_par}. 
The joint analysis yields a column density of 
$n_{H}=(2.5^{+1.0}_{-0.7})\times10^{21}$~cm$^{-2}$.  
The photon index and the normalization at 1~keV for the pulsar are 
found to be $\Gamma_{X}=2.2^{+0.2}_{-0.3}$ and 
$(8.3^{+2.1}_{-1.6})\times10^{-6}$~photons~keV$^{-1}$~cm$^{-2}$~s$^{-1}$ 
respectively. The corresponding best-fit spectral 
parameters for the jet are $\Gamma_{X}=1.2\pm0.1$ and 
$(8.7^{+1.0}_{-1.5})\times10^{-6}$~photons~keV$^{-1}$~cm$^{-2}$~s$^{-1}$ 
respectively. In the energy band of $0.5-10$~keV, 
the observed flux and absorption-correct fluxes for \psr\ are 
found to be $(2.4^{+1.8}_{-1.0})\times10^{-14}$~ergs~cm$^{-2}$~s$^{-1}$
and $(3.4^{+1.7}_{-1.0})\times10^{-14}$~ergs~cm$^{-2}$~s$^{-1}$ respectively. 
And the observed flux and absorption-corrected flux 
for its jet in the same energy band are 
$(9.1^{+3.2}_{-2.8})\times10^{-14}$~ergs~cm$^{-2}$~s$^{-1}$ and 
$(10.0^{+3.1}_{-2.7})\times10^{-14}$~ergs~cm$^{-2}$~s$^{-1}$ respectively. 

X-ray emission of the jet can possibly be powered by the energy loss of the synchrotron emitting leptons during their rides 
from the pulsar. If the synchrotron cooling time is less than the timescale for the flow of leptons across the 
feature, steepening of the photon index for the synchrotron radiation is expected along the jet. 
In order to investigate this possible spectral variation along the jet, we divide the extraction region 
for the whole jet equally into two parts along its orientation. In the following, we refer the segment closer to the pulsar 
as region~1 and the further one as region~2. The total photon statistics collected 
by three EPIC cameras for region 1 and region 2 are $379\pm28$~cts and $376\pm26$~cts respectively. 
For modeling the spectra of these segments, we also consider the power-law model. For a more 
constraining analysis, we again jointly fit the spectra of regions 1 and 2 as well as the whole jet with the column density      
in the individual model tied together. The results are summarized in Table~\ref{spec_par}. 
In this joint analysis, the X-ray 
spectra in both regions can be well-modeled with a power-law with a slope $\Gamma_{X}\sim1.3$ and hence no variation of 
photon index have been found. 

Since a knot-like X-ray feature along the jet is identified and a possible optical source is found within its 
positional uncertainty, this leads us to speculate whether this is a field object just happens to locate in the error circle 
by chance instead of intrinsically related to the jet. Therefore, we also attempt to examine the X-ray spectrum of this 
knot-like feature. Within a circular region with a radius of $15''$ centered at its nominal position (see \S3.1), 
there are $100\pm15$~cts collected by all three EPIC cameras after background subtraction. We have examined its spectrum 
by fitting with various single component models which include a power-law, a thermal bremsstrahlung as well as a 
collisional ionization equilibrium plasma model. In view of the limited photon statistic of this feature, 
we simply assume the column absorption is at the level constrained in the aforementioned joint analysis 
(i.e. $n_{H}=2.8\times10^{21}$~cm$^{-2}$). We found that all three tested models result in a similar 
goodness-of-fit ($\chi^{2}=18.1$ with 23 d.o.f.) and hence we are not able to discriminate these competing 
scenarios. We notice that all the models yield an observed flux at the level of
$f_{X}\sim1.5\times10^{-14}$~ergs~cm$^{-2}$~s$^{-1}$ ($0.5-10$~keV). In the HST image, the possible 
optical counterpart of this feature has a magnitude of $R\sim22.31$. This gives rise to an X-ray-to-optical 
flux ratio of $f_{X}/f_{R}\sim0.77$. 
This value is higher than that expected for a field star which typically has a ratio of $f_{X}/f_{\rm opt}<0.3$
(Maccacaro et al. 1988) but conforms with the expected range of an AGN which typically has a ratio of 
$f_{X}/f_{\rm opt}<50$ (Stocke et al. 1991). 

Since the nature of the knot-like feature cannot be confirmed, we have also reexamined the spectral behavior 
of the jet with the knot removed as if it were a background source. In particular, as the knot is located in region 1, 
we would like to investigate whether its removal has an effect in the spectral shape of this region. 
After subtracting the contribution from the knot, we re-do the joint spectral fitting of the spectra of 
regions 1 and 2 as well as the whole jet. The results are also summarized in Table~\ref{spec_par}. 
The photon indices in these two regions remain to be consistent within $1\sigma$ uncertainties. Therefore, 
we conclude that no evidence for spectral steepening along the X-ray jet can be found in this observation. 

\subsection{Temporal Analysis}

\subsubsection{Search for the coherent pulsation from \psr}

In order to search for X-ray pulsations from the pulsar PSR B2224+65,
only the data taken from the PN small window mode with a temporal
resolution of 5.7 ms can be used. We extracted 1450 counts from a circle
with a radius of $20''$ centered at the radio timing position. We further
selected these events within the good time intervals with the counts rate
of the entire field 
lower than 0.02 cts/sec in order to reduce the contamination from the sky
background. After filtering, a total of 523 source counts was left in our
study. The photon arrival times were corrected to the solar system barycenter
with the BARYCEN tool (version: 1.18; JPL DE200 Earth ephemeris) of XMMSAS. 

We noticed that the pulsar has a glitch at MJD~$\sim$54266, which is 
about two years before our observation (MJD~55025). Since there is no 
further information about the relaxation time, we then performed a detailed 
search for a range that cover both the post-glitch frequency as well as the 
extrapolated value at the mean epoch of the XMM-Newton observation. 
We applied the $Z^{2}_{n}$ statistics (Buccheri et al.~1983) with the 
harmonics number (n) from one to ten and the H-Test (de Jager, 
Raubenheimer \& Swanepoel 1989) with the searching step of 1/5 corresponding Fourier width
in the range of 1.4650 to 1.4652~s$^{-1}$. No significant signal was detected.
We also folded the light curve by using the radio spin period extrapolated for 
the epoch of the XMM-Newton observation. However, no meaningful light curve 
was obtained. The pulsar ephemeris reported from the ATNF Catalogue (Manchester et al. 2005), 
$f = 1.46511023680$~Hz and $\dot{f} = -2.0737 \times 10^{-14}~\mbox{sec}^{-2}$ 
(at MJD = 54420.0) were used in our study. 
We further computed the upper limit of the pulsed fraction $f_{p}$ by 
$f_{p}=\beta\left(N_{t}-N_{b}\right)/N_{t}$, where $N_{t}$ and $N_{b}$ are 
the number of total photons and the background photons respectively 
and $\beta$ is the duty cycle. Assuming $\beta=0.5$, we placed a $1\sigma$ 
upper limit for the pulsed fraction to be $\sim8\%$. 

\subsubsection{Search for long-term variability from the jet}

As an X-ray jet is possibly subjected to magnetohydrodynamic instability, variability has been seen 
in many PWN systems (e.g. DeLaney et al. 2006). In order to investigate the long-term 
variability from the jet, we analyse the X-ray observations of this field in different epochs.  
We have ignored one of the Chandra observations (Obs ID. 6691) as its short exposure 
($\sim10$~ks) does not provide sufficient photon statistic for this analysis. The multi-epoch 
spectral results are summarized in Table~\ref{jet_var}. Since the flux estimates are sensitively 
to the column absorption, we fixed the $n_{H}$ in all observations at the value inferred from 
the analysis of the XMM-Newton observation as it provides a superior photon 
statistic in the soft band and hence put a better estimate on the $n_{H}$. Also, for the sake comparison, 
we compute the flux in an energy range of $0.5-10$~keV for all cases. We found that both 
observed and absorption-corrected fluxes, as well as  the photon index are consistent among the observations 
in different epochs within 
$1\sigma$ uncertainties. Therefore, we do not find any evidence of the long-term spectral and flux 
variability. 

Apart from searching for the long-term variability, we have also investigated if there is any noticeable 
flux variation within each individual exposure. By examining the light curves, we also do not find any 
evidence for the variability within each observation window. 

\section{SEARCH FOR $\gamma-$RAY EMISSION WITH FERMI LARGE AREA TELESCOPE}

\psr\ has a spin-down age of $\sim10^{6}$~yrs. 
It has been speculated that the accelerating regions can still be sustained in the magnetospheres of some old pulsars 
(e.g. Zhang et al. 2004; Hui \& Becker 2007). Indeed, pulsed $\gamma-$rays with energies $>0.1$~GeV 
have recently been detected from an old radio-quiet pulsar, PSR~J1836+5925, 
that has a characteristic age of 1.8 million years 
(Abdo et al. 2010). Motivated by this discovery, we also attempt to search for the $\gamma-$ray emission from \psr. 

The continuous release of the photon data obtain in the all-sky survey with the Large Area Telescope (LAT) on board the Fermi 
Gamma-ray Space Telescope enables us to search for the $\gamma-$ray emission from \psr\ in MeV$-$GeV regime. 
In this analysis, we used the data obtained between 2008 August 4 and 2011 Oct 2. We used the Fermi Science Tools v9r23p1 
package to reduce and analyze the data in the vicinity of \psr. Only events that are classified as the 
``Source'' class\footnote{An event class in Pass 7 (i.e. evclass=2) that provides the most suitable photon data for point 
source spectral analysis.} were used. 
To reduce the contamination from Earth albedo $\gamma$-rays, we excluded events with zenith angles greater than 100$^\circ$. 
The instrument response functions ``P7SOURCE\_V6'' were used. Events with energies between 100~MeV and 20~GeV were 
used for our analysis. 

With the aid of the task \emph{gtlike}, we carried out an unbinned maximum-likelihood analysis for a region-of-interest (ROI)
with a 5$^\circ$ radius centered on the position of \psr\ in the energy range of $0.1-20$~GeV. 
We subtracted the background contribution 
by including the Galactic diffuse model (gal\_2yearp7v6\_v0.fit) and the isotropic background (iso\_p7v6source.txt), as well as 
all sources in the second Fermi/LAT catalog (2FGL; Abdo et al. 2011) within $10^\circ$ from \psr.  
We assumed a power-law spectrum for \psr\ as well as all 2FGL sources in our consideration. 
Making use of the model parameters resulted from the maximum-likelihood analysis, we computed the test-statistic (TS) map of 
$2^{\circ}\times2^{\circ}$ centered at the nominal position of \psr. The 
TS value\footnote{TS value corresponds to the detection significance of $\sigma\approx\sqrt{\rm TS}$.}
at the position of \psr\ is found to be $\sim40$. 
However, we speculated that this can possibly be contributed 
by two previously unknown nearby $\gamma-$ray sources with their TS values of $\sim29$ and $\sim38$ 
peaked at RA=336.82$^\circ$ Dec=65.95$^\circ$ (J2000) 
and RA=335.25$^\circ$ Dec=64.98$^\circ$ (J2000) respectively. Appending these two additional sources in the model,  
we have repeated the maximum-likelihood analysis and the TS map computation.  
With the contribution from these two serendipitous sources subtracted, the TS value at the position of \psr\ essentially drops 
to zero and no residual emission can be identified in the TS map. 
Therefore, we conclude that we do not find any evidence for $\gamma-$ray emission from \psr\ 
in this study. Adopting a power-law photon index of $\Gamma=2$, we place a $1\sigma$ upper-limit of 
$\sim4\times10^{-12}$~erg~cm$^{-2}$~s$^{-1}$ in $0.1-20$~GeV.  

\section{DISCUSSION}

\subsection{Nature of the X-ray emission from \psr\ \& its jet}

We have presented a detailed investigation of the X-ray emission from \psr\ and its jet with XMM-Newton. 
The non-detection of any coherent X-ray pulsation suggests that the magnetospheric 
origin is questionable. 
In the context of outer-gap model, the non-thermal pulsed X-rays are resulted from 
the back-flowing charge particles from the outer-gap (e.g. Cheng \& Zhang 1999). Based on this model, the 
electromagnetic cascade developed in this flow should give rise to an X-ray photon index $\leq2$. 
The photon index of the X-rays from \psr\ is $2.2^{+0.2}_{-0.3}$. The best-fit value is found to larger than 
that expected from the magnetospheric model, though they can be reconciled within $1\sigma$ uncertainties. 
 
On the other hand, with the deep X-ray imaging of \psr, we have revealed an interesting feature at the pulsar position which 
is elongated along direction of proper motion as observed in different cameras (cf. insets in Fig.~\ref{xmm_img}). 
Although we cannot confirm its extent unambiguously 
with the existing data, its orientation is interesting and worth discussing. X-ray tails have been found to be associated 
with many pulsars (see Kargaltsev \& Pavlov 2010 for a recent review). 
The tails are usually oriented in the opposite direction of pulsar motion and behind the bow-shock, 
therefore it is interpreted as the flow of particles coming out of the ram-pressure dominated PWN. In this 
scenario, the X-ray emission from \psr\ is originated from the synchrotron radiation from the wind particles 
accelerated at the termination 
shock. This is consistent with the unpulsed non-thermal emission inferred from this observation. For the synchrotron emitting 
leptons of Lorentz factor $\gamma$ distribute as $N\left(\gamma\right)\propto\gamma^{-p}$, the observed photon index 
of $\Gamma\sim2.2$ from the pulsar location is consistent with the value expected from the fast-cooling scenario, 
$\Gamma=\left(p+2\right)/2$, for the electron index lies within a typical range of $p\sim2-3$ (Cheng, Taam, \& Wang 2006).

This putative feature, which has an extent of $\sim8''$ in EPIC images, should be resolved by Chandra. 
Nevertheless, in reexamining the archival Chandra data, its extent cannot confirm. On the other hand, the photon 
distribution of \psr\ in the merged Chandra image is slightly deviated from a point source with the deformation also 
toward the direction behind the pulsar proper motion. Although the evidence for the extended source cannot be 
conclusive with the existing data, the compact feature suggested by the Chandra data can possibly be the 
portion of the highest surface brightness. Complex structures of bow-shock PWNe have been observed in other 
systems. For example, XMM-Newton observation of a nearby pulsar PSR~B1929+10 has revealed a long X-ray tail  
of several arc-minutes (Becker et al. 2006). Apart from confirming the detection 
of this long trail, a follow-up Chandra observation has further resolved a brighter but more compact feature 
with an extent of $\sim10''$ around the pulsar (Hui \& Becker 2008). 
For a further investigation of \psr, a dedicated deep Chandra observation, which simultaneously
provides a sub-arcsecond spatial resolution and a photon statistic comparable with this XMM-Newton observation,
is the only way to confirm or refute this suggested feature.

While this compact feature is still questionable, the evidence for the $\sim2'$ long jet is unambiguous. Although 
all the X-ray studies suggest its non-thermal nature, its physical origin remains to be obscure. Very recently,
a collimated $\sim9'$ long X-ray feature has been discovered from a radio-quiet $\gamma-$ray pulsar PSR~J0357+3205 
(De Luca et al. 2011) which is very similar to the jet of \psr. 
It is interesting to compare the observed properties between these two systems. 
The spin-down luminosity of PSR~J0357+3205, $\dot{E}\sim6\times10^{33}$~erg/s, is $\sim6$ times higher than that of \psr, 
but it is the lowest among all known non-recycled $\gamma-$ray pulsars. The low column absorption suggests it is a nearby 
pulsar with a distance of only a few hundred parsec. Assuming there is no inclination with respect to the sky plane, 
this suggests the physical extent of the trail associated with PSR~J0357+3205 at the order 
of $\sim1$~pc which is similar to the case of \psr. In both cases, there is no evidence of spectral variation along 
the feature. For a distance of 500 pc, the X-ray conversion efficiency $L_{x}/\dot{E}$ 
of its trail in $0.5-10$~keV is $\sim10^{-3}$. Adopting a distance of 1~kpc for \psr, $L_{x}/\dot{E}$ of its jet in the same energy 
band is found to be $\sim10^{-2}$ which is an order larger than that for PSR~J0357+3205. This might be why the jet of \psr\ can 
still be detected in spite of its larger distance. 

Following De Luca et al. (2011), we estimate the highest achievable energy of synchrotron-radiating particles 
injected by \psr. First, we assume the highest energy of the electrons/positrons can be gained is comparable with maximum potential drop 
in the magnetosphere, which is $\Delta\Phi=\left(3\dot{E}/2c\right)^{1/2}$ for an aligned pulsar (Goldreich \& Julian 1969). Similar 
to the case of PSR~J0357+3205, this results in a maximum Lorentz factor of $\gamma_{\rm max}\sim10^{8}$. In the presence of a 
magnetic field, these particles will radiate synchrotron emission with characteristic frequency of $\nu_x=\gamma^2eB/m_ec$, where 
$B_{\mu G}$ is the local magnetic field strength in the emission region in units of microgauss. Adopted a typical 
magnetic field at an order of $\sim5~\mu$G, $\gamma_{\rm max}$ implies a synchrotron 
photon energy at the order of $h\nu_x\sim5$~keV. 

One difficulty for the synchrotron scenario is that it requires the Lorentz factor of the emitting particles to achieve the 
same order of $\gamma_{\rm max}$ so as to produce X-rays at a few keV. On the other hand, inverse Compton (IC) scattering between 
the relativistic particles and the relic photons appears to be more viable in producing the observed X-rays. 
We note that the energy density of the relic photons is at the order of $\sim10^{-12}$~erg~cm$^{-3}$, which is 
comparable with the magnetic energy density for a field strength of $\sim5\mu$G. 
Therefore the IC power cannot be ignored. In the IC scenario, the required 
Lorentz factor to produce $\sim10$~keV photons is only at the order of $\gamma\sim10^{4}$ which is much easier to achieve. 

The major difficulty for explaining the X-ray jet of \psr\ in the context of PWN comes from its orientation with 
respect to the pulsar's proper motion. The jet protrudes from the pulsar and extends toward northwest, which makes an angle 
of $\sim118^{\circ}$ from the proper motion direction of the pulsar (cf. Fig.~\ref{xmm_img}). 
This peculiar orientation makes the bow-shock interpretation 
questionable as this scenario requires the nebular emission lies in an opposite direction of the pulsar's velocity vector. 
In view of this difficulty, different explanations for the nature of the jet association with \psr\ have been proposed, 
ranging from the magnetically confinement of relativistic particles leaking from the bow-shock (Bandiera 2008) 
to a picture similar to the AGN jet outflow (Johnson \& Wang 2010). Nevertheless, there is no consensus on its physical origin yet. 

Since the extended feature of PSR~J0357+3205 is akin to that of \psr, a comparative analysis between these two systems 
might provide us a deeper insight on the nature of their jets. While the distance and the proper motion direction of \psr\ 
are well-constrained, these are the most important parameters to be determined in further investigation of 
PSR~J0357+3205. These would enable a more reliable estimations
of its energetic and most importantly the orientation of the jet with respect to the pulsar's space velocity, which will 
allow us to ascertain if these two systems have a similar physical origin. 

\subsection{The origin of \psr}
\psr\ is certainly one of the most spectacular neutron stars regarding its motion through space as 
its transverse velocity, $v=865(d/1~{\rm kpc})$~km~s$^{-1}$, is among the highest in the currently known pulsar population. 
On the other hand, the well measured bow shock with its inclination in the plane of the sky 
suggests a negligible radial velocity component (Chatterjee \& Cordes 2004) 
which has been confirmed recently by Tetzlaff et al. (2009) who suggest the origin of \psr\ to lie within the Cygnus OB3 association 
or a small massive cluster nearby. This origin indicates a small radial velocity of $v_{r}=-21^{+81}_{-70}$~km~s$^{-1}$. 
The proposed kinematic age of $\tau_{\rm kin}\approx 0.8$ Myr is consistent with a characteristic pulsar age of 
$\tau_{\rm char}=1.1$~Myr (Hobbs et al. 2004; Manchester et al. 2005). 

To advance the analysis we presented earlier (Tetzlaff et al. 2009), we attempt to identify a runaway star which might 
have been the former companion of this pulsar's progenitor star (we refer to such runaway stars as 
binary supernova scenario (BSS) runaways; Blaauw 1961). Therefore, we take all runaway star candidates with full kinematics 
from the runaway star catalogue of Tetzlaff et al.
(2011) and compare the past flight paths of \psr\ and
each runaway star varying the observables by performing a Monte-Carlo simulation to find close encounters 
between the pulsar and the runaway star. Simultaneously, we calculate the distance to possible 
parent associations/clusters listed in Tetzlaff et al. (2010) 
(for a full description of the procedure we refer to Tetzlaff et al. 2009 and Tetzlaff et al.
2010).\footnote{\psr\ reaches a distance to the Sun larger than 3 kpc (the catalogue limit of the current runaway positions) 
if it is older than 1.3 Myr (assuming $v_{r}=0$~km/s). Since the characteristic age ($\sim1$ Myr) provides an upper limit on the true 
age of young (few Myr) pulsars in most cases, it is well possible to identify the associated runaway star (the former companion, 
if it exists) among the catalogue runaway star candidates.}

For \psr\ we take the following data for the equatorial coordinates, right ascension $\alpha$ and 
declination $\delta$, the distance to the Sun $d$ and the proper motion 
components $\mu_{\alpha}^*$ ($\mu_{\alpha}^{*}=\mu_{\alpha}\cos\delta$) and $\mu_{\delta}$ 
(Harrison et al. 1993; Taylor \& Cordes 1993; Yuan et al. 2010); 

\begin{equation}
        \begin{array}{l c l}
        \alpha &=& 336^\circ\hspace{-0.8ex}.469671,\ \delta\ =\ 65^\circ\hspace{-0.8ex}.593217,\\
        d &=& 2\pm1~{\rm kpc},\\
        \mu_{\alpha}^* &=& 144\pm3~{\rm mas~yr}^{-1},\ \mu_{\delta} \ =\ 112\pm3~{\rm mas~yr}^{-1}.
        \end{array}
\end{equation}

Here we adopt $d=2$~kpc for the calculation as published by Taylor \& Cordes (1993) and 
adopt a reasonable error of 1~kpc that covers the distance inferred by Cordes \& Lazio (2002) and 
Chatterjee \& Cordes (2004).
Among 1703 runaway star candidates in the catalogue with full kinematic data, 
only nine appear to show possible close encounters with \psr\ 
(HIP 97359, 98773, 99283, 99580, 100308, 100409, 101186, 107418, 109393). 
For two of them, HIP 107418 and HIP 109393, separations between the runaway star 
and the pulsar are small at $\tau_{\rm kin}\approx0$, i.e. 
at present; thus we exclude these two stars from our analysis.

Tracing back the seven remaining runaway stars and OB associations/
clusters listed in Tetzlaff et al. (2010), we find that four stars
could have been ejected from their parent cluster within the last
3 Myr ($\approx3\cdot\tau_{\rm char}$) (see Table~\ref{runaway}). 
For them, we simultaneously trace back \psr, the runaway star and the 
potential parent association varying the observables within their confidence intervals 
(here we adopt $v_r=0\pm100$~km s$^{-1}$ for \psr) to find the closest approach 
between the pulsar and the runaway star (i.e. the minimum separation $d_{\rm min}\left(\tau_{\rm kin}\right)$)
as well as the associated time ($\tau_{\rm kin}$) and the distance to the respective association centre 
at that time.

In only one case -- HIP 99580 (=V2011 Cyg) with Cyg OB9 being the parent association -- 
the pulsar and the runaway star can be located within the association boundaries 
(radius of Cyg OB9 $\approx40$~pc) at the same time in the past. The smallest separation 
$d_{\rm min}$ between \psr\ and HIP 99580 that is found after 15 million Monte-Carlo runs 
is $0.4$~pc. Between 25~pc and 35~pc distance from the centre of Cyg OB9
\footnote{We choose a distance range of 10~pc to make it comparable to the previously suggested neutron star/ 
runaway pair PSR B1929+10/$\zeta$ Ophiuchi, see Hoogerwerf et al. (2001) and Tetzlaff et al. (2010)}, 
we find a peak of small separations $d_{\rm min}$ between the pulsar and the runaway star. 
In 2540 simulations the distance of the pulsar as well as the distance of the runaway star 
to the centre of Cyg OB9 falls within this range. The first part of the distribution of the separations 
$d_{\rm min}$ between \psr\ and HIP 99580 of these runs 
is consistent with the hypothesis that both stars once were at 
the same position $0.65\pm0.05$~Myr in the past (see Figure~\ref{histo}). 

Although only a small fraction of runs yield small separations $d_{min}$ between 
the stars as well as the centre of Cyg OB9 the scenario seems likely considering 
the errors on the parameters involved (distance $d$, proper motion components 
$\mu_\alpha^\ast=\mu_\alpha\cos\delta$ and $\mu_\delta$, radial velocity $v_r$ 
of both stars\footnote{The errors on the motion of Cyg OB9 are also taken into 
account but are negligible compared to the stars.}). The fraction of successful runs 
is much smaller than in the case of the known neutron star/ runaway pair PSR B1929+10/ 
$\zeta$ Ophiuchi (about one order of magnitude; for PSR B1929+10/ $\zeta$ Oph three million 
Monte-Carlo runs were performed approximately twice the number of successful runs compared 
to 15 million here; see Hoogerwerf et al. 2001 and Tetzlaff et al. 2010); 
however, since \psr\ and HIP 99580 are far more distant from the Sun than PSR~B1929+10 and 
$\zeta$ Oph (the error on the distance is larger by about one to two orders of magnitude), this is to be expected.

In Figure~\ref{correlation}, we show the correlations between the astrometric parameters 
which were used as input for the 2540 runs for which both, \psr\ and the 
runaway star HIP 99580, were situated between 25~pc and 35~pc  
from the centre of Cyg OB9. For the runaway star only little correlation between 
$\pi$ and $v_r$ can be seen and the most probable values of all parameters fall clearly 
within the error bars. Whereas for the pulsar there is a strong correlation
between $\pi$ and $v_r$. This is expected since an increase in distance (decrease in $\pi$) 
can be compensated for by a larger radial velocity $v_r$ and vice versa. 
If \psr\ and the runaway star HIP 99580 were at the same place 
0.65~Myr ago in Cyg OB9, i.e. once were in the same multiple system, 
the present distance of \psr\ needs to be $\sim0.8$~kpc. 
Hence, we can confirm previous findings by Chatterjee \& Codes (2004) which concludes 
the pulsar have a distance to the Sun of $\sim1$~kpc. 

If this supernova occurred $0.65$~Myr in the past, it was located at Galactic coordinates 
$l=77.4\pm0.2~{\rm deg}, b=2.5^{+0.4}_{-0.2}~{\rm deg}$\footnote{These coordinates are as viewed 
at the present time. (i.e. Without the correction for the Sun's motion through the Galaxy over the 
last $0.65$~Myr.)},  
and a distance to Earth of $d_{\rm Earth}=968^{+32}_{-28}$~pc. 

If HIP 99580 is a BSS runaway star, it should show signs of stellar evolution in the former system such 
as a large rotational velocity and an enhanced helium abundance caused by momentum and mass transfer during 
evolution. It was claimed by Barannikov (1993) that HIP 99580 is a spectroscopic binary, however 
De Becker \& Rauw (2004)\footnote{We note that these authors quote a different radial velocity of
$v_r=-23.1\pm8.0$~km~s$^{-1}$ for HIP 99580 compared to $-44\pm4$ km/s (Kharchenko et al. 2007) that we used here.
In this case, the star would
still be a runaway star as it was identified also from its tangential velocity
(Tetzlaff et al. 2011). However, only a tiny fraction of runs would then yield small
separations between HIP 99580 and \psr\ making the BSS scenario unlikely. But note that
both radial velocities can be correct if the star is a spectroscopic binary, then it
would be necessary to determine the systematic radial velocity.}  
and Turner et al. (2008) could not confirm this. According to 
Conti \& Ebbets (1977), the O5 type star HIP 99580 is a rapid rotator with $v\sin{i}=270$~km~s$^{-1}$. Penny (1996) 
investigated $v\sin{i}$ of O-type
stars and found that only a small number of very massive stars show such a large $v\sin{i}$. 
Hence, it is reasonable to assume that it gained its large $v\sin{i}$ during the earlier evolution in a binary system.
Furthermore, BSS runaway stars should be blue stragglers, i.e. they appear younger, hence bluer, 
than their parent association (see Hoogerwerf et al. 2001 for other examples). 
In Figure~\ref{bluestraggler} we show the colour-magnitude diagram of the potential parent association Cyg OB9 with the data adopted 
from Garmany \& Stencel (1992). Apparently, the association consists of two major groups of different ages, the younger 
one being $\sim4$ Myr old.\footnote{\bf Note that Uyan{\i}ker et al. (2001) quote an age for Cyg OB9 of 8~Myr.} 
If HIP 99580 originated from Cyg OB9, it is clearly a blue straggler. 

Assuming an association age of 4~Myr, the progenitor star of the neutron star
must have been 3.4~Myr old as it exploded. Thus, its mass was $(67\pm15)~M_{\odot}$
(Tinsley 1980; Maeder \& Meynet 1989; Kodama 1997). For such large masses, formation of black holes 
is expected rather than neutron stars (upper limit $\approx30~M_{\odot}$, Heger et al. 2003). However, due to mass transfer 
during binary evolution, even more massive stars ($\gtrsim50-60~M_{\odot}$) may eventually
produce neutron stars instead of black holes (Heger et al. 2003; Belczynski \& Taam 2008). Also, more massive 
progenitors may also produce larger speeds of the neutron star (Burrows \& Young 2000) which is consistent with 
the extraordinary speed of \psr.

\acknowledgments
The authors would like to thank Ralph Neuh\"{a}user for reading the manuscript and
provide useful comments for improving its quality. CYH and LT would like to thank Albert Kong and 
National Tsing Hua University for the hospitality during their visit in 2010 summer. 
CYH is supported by the National Research Foundation of Korea through grant 2011-0023383. 
NT acknowledges financial support from Carl-Zeiss-Stiftung and partial support from Deutsche Forschungsgemeinschaft 
(DFG) in the SFB/TR-7 Gravitational Wave Astronomy. LT would like to thank DFG for financial support in project SFB TR7 Gravitational 
Wave Astronomy. KSC is supported by the GRF Grants of
the Government of the Hong Kong SAR under HKU 7009/11P.

\clearpage

\clearpage

\clearpage

\clearpage

\clearpage
\begin{figure*}
\centerline{\psfig{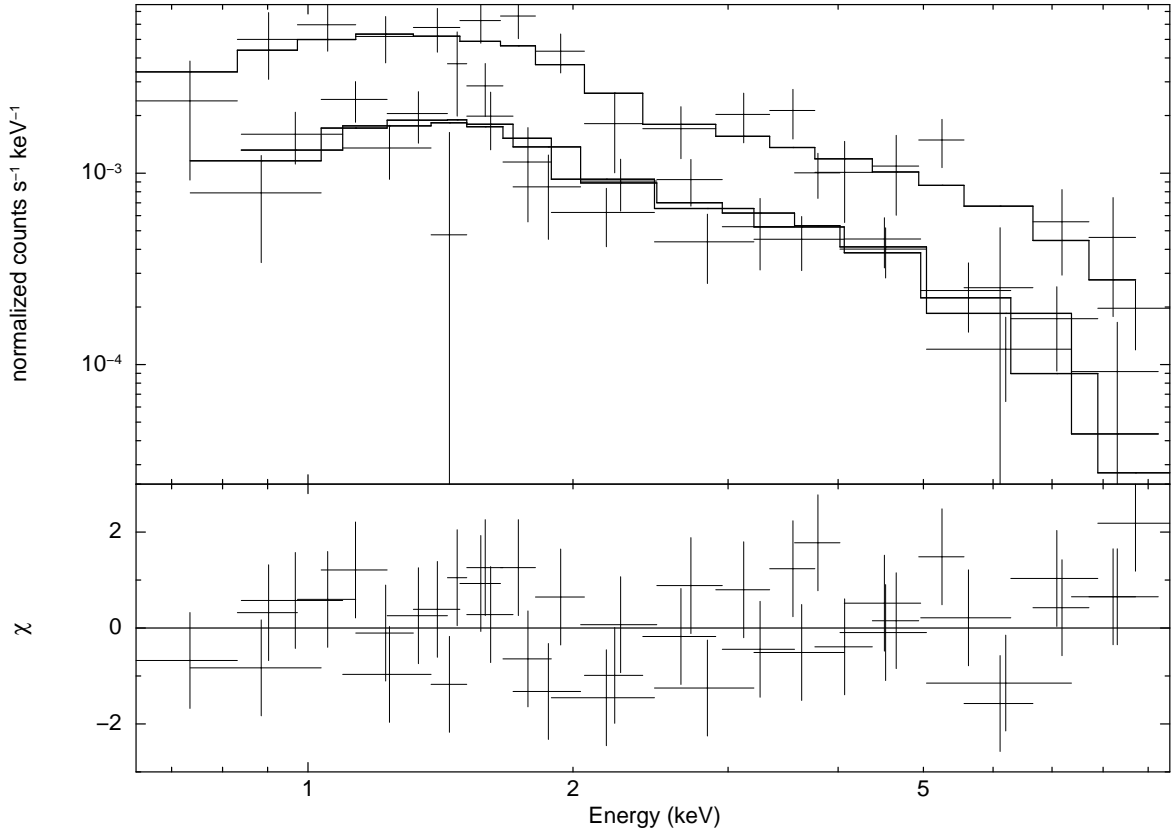}}
\caption[]{X-ray spectrum of the entire long jet feature associated with \psr\  
as observed with the PN (upper spectrum) and MOS1/2 detectors (lower spectra) and 
simultaneously fitted to an absorbed power-law model ({\it upper panel}) and contributions to the $\chi^{2}$ fit statistic 
({\it lower panel}).}
\label{jet_spec}
\end{figure*}

\clearpage
\begin{center}
\begin{deluxetable}{lcccccc}
\tablewidth{0pc}
\tablecaption{X-ray spectral parameters of \psr\ and its jet.\tablenotemark{a}}
\startdata
\hline\hline
Segment & $n_{H}$ & $\Gamma_{X}$ & $f^{\rm obs}_{\rm 0.5-10~keV}$\tablenotemark{b} & $f^{\rm unabs}_{\rm 0.5-10~keV}$\tablenotemark{b} & $\chi^{2}$ & d.o.f \\
        & $10^{21}$~cm$^{-2}$ & & $10^{-14}$~ergs~cm$^{-2}$~s$^{-1}$ & $10^{-14}$~ergs~cm$^{-2}$~s$^{-1}$ & & \\\hline
\multicolumn{7}{c}{Pulsar + Whole jet}\\\hline
Pulsar & $2.5^{+1.0}_{-0.7}$ & $2.2^{+0.2}_{-0.3}$ & $2.4^{+1.8}_{-1.0}$  & $3.4^{+1.7}_{-1.0}$ & 42.2 & 49 \\
Whole jet   & ... & $1.2\pm0.1$ & $9.1^{+3.2}_{-2.8}$ & $10.0^{+3.1}_{-2.7}$ & ... & ... \\\hline
\multicolumn{7}{c}{Whole jet + Region 1 + Region 2}\\\hline
Whole jet  & $2.8^{+0.9}_{-0.6}$ & $1.3\pm0.1$ & $8.9^{+3.8}_{-2.5}$ & $10.0^{+3.8}_{-2.5}$ & 86.2 & 91 \\
Region 1   & ... & $1.3\pm0.1$ & $4.4^{+1.4}_{-1.5}$ & $5.0^{+1.4}_{-1.6}$ & ... & ... \\
Region 2   & ... & $1.3^{+0.2}_{-0.1}$ & $4.1^{+1.6}_{-1.9}$ & $4.6\pm2.0$ & ... & ... \\\hline
\multicolumn{7}{c}{Whole jet + Region 1 + Region 2 (with the knot removed) }\\\hline
Whole jet  & $3.2^{+1.0}_{-0.8}$ & $1.1\pm0.2$ & $6.9^{+4.5}_{-3.1}$ & $7.7^{+4.6}_{-3.1}$ & 69.2 & 79 \\
Region 1   & ... & $1.1\pm0.2$ & $4.3^{+2.9}_{-1.7}$ & $4.7^{+2.9}_{-1.7}$ & ... & ... \\
Region 2   & ... & $1.3^{+0.2}_{-0.1}$ & $4.6^{+2.3}_{-1.3}$ & $4.6^{+2.3}_{-1.3}$ & ... & ... \\
\enddata
\tablenotetext{a}{Following the method adopted by Johnson \& Wang (2010), these parameters are obtained from the 
joint analysis of the spectra of segments of interest with the column densities in the individual models tied together.}
\tablenotetext{b}{$f^{\rm obs}_{\rm 0.5-10~keV}$ and $f^{\rm unabs}_{\rm 0.5-10~keV}$ represent the observed and absorption-corrected fluxes in $0.5-10$~keV respectively.}
\label{spec_par}
\end{deluxetable}
\end{center}

\clearpage
\begin{center}
\begin{deluxetable}{lccccc}
\tablewidth{0pc}
\tablecaption{X-ray spectral properties and the fluxes of the jet-like feature at different epochs.}
\startdata
\hline\hline
Start date & Obs. ID. & n$_H$             & $\Gamma_{X}$ & $f^{\rm obs}_{\rm 0.5-10~keV}$ &  $f^{\rm unabs}_{\rm 0.5-10~keV}$ \\
             &             & 10$^{21}$ cm$^{-2}$ &   & 10$^{-14}$ ergs~cm$^{-2}$~s$^{-1}$   & 10$^{-14}$ ergs~cm$^{-2}$~s$^{-1}$\\\hline
\multicolumn{6}{c}{Chandra Observations} \\
\hline
2000-10-21 & 755 & 2.7 (fixed)  & $1.1\pm0.4$   & $8.7^{+16.1}_{-5.5}$ &9.5$^{+16.2}_{-5.8}$  \\
2006-10-06 & 7400  & 2.7 (fixed)  & 1.0$\pm0.4$ & $8.7^{+11.3}_{-5.0}$ &9.4$^{+11.6}_{-5.3}$ \\
\hline
\multicolumn{6}{c}{XMM-Newton Observation} \\
\hline
2009-07-13 & 0604420101 & $2.7^{+1.2}_{-0.8}$  &  $1.2\pm0.1$  & $9.0^{+4.7}_{-3.0}$ & 10.0$^{+4.7}_{-3.0}$ \\
\enddata
\label{jet_var}
\end{deluxetable}
\end{center}

\clearpage
\begin{center}
\begin{deluxetable}{lccccc}
\tablewidth{0pc}
\tablecaption{Runaway star candidates with possible close encounters to the Guitar pulsar and a possible kinematic age 
$\tau_{\rm kin,*}\lesssim3$~Myr ($\approx3\cdot\tau_{\rm char}$). The possible parent association and the associated 
$\tau_{\rm kin,*}$ are given in Columns 2 and 3. Columns 4 and 5 quote the star's age $\tau_{*}$ and mass $M_{*}$ as inferred from 
evolutionary models (see Tetzlaff et al. 2011). The last Column indicates the spectral type as given in Tetzlaff et al. (2011).}
\startdata
\hline\hline
HIP & parent  & $\tau_{kin,*}$ & $\tau_{*}$ &  $M_{*}$ & SpT\\     
    &         &  Myr          & Myr         & $M_\odot$ & \\\hline
98773  & Cyg OB7 & $3.0^{+1.0}_{-1.0}$ & $11.6\pm0.5$ & $15.5\pm0.8$ & A2V+...\\
99580  & Cyg OB9\tablenotemark{a} & $1.0^{+1.0}_{-1.0}$ & $1.5\pm0.5$\tablenotemark{b} & $44.6\pm5.1$ & O5e \\
100409 & Cyg OB9 & $3.6^{+1.4}_{-0.6}$ & $11.1\pm1.1$ & $15.5\pm2.7$ & B1Ib\\
       & Cyg OB7 & $5.4^{+1.6}_{-2.4}$ & ... & ... &  ...   \\
       & Cep OB6 & $4.0^{+1.0}_{-1.0}$ & ... & ... &  ...   \\
101186 & Cyg OB9\tablenotemark{a} & $2.1^{+0.8}_{-1.2}$ & $3.9\pm0.1$ & $37.2\rm0.1$ & O9.5Ia
\enddata
\tablenotetext{a}{Note that Schilbach \& R\"oser (2008) suggested an origin for HIP 99580 
and HIP 101186 in the cluster NGC~6913 2.4~Myr and 2.9~Myr ago, respectively. 
This cluster lies within the Cyg OB1/ Cyg OB9 region.}
\tablenotetext{b}{This age does not take into account that the star might be a blue straggler. 
As blue straggler, its age can be consistent with 4 Myr (cf. Fig. 9).}
\label{runaway}
\end{deluxetable}
\end{center}

\clearpage
\begin{figure*}
\centerline{\psfig{figure=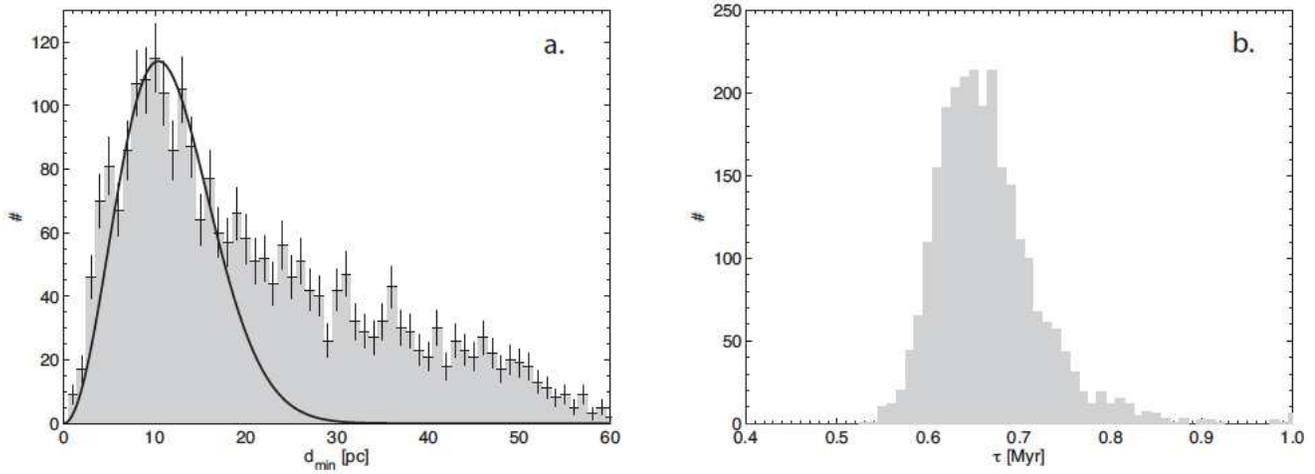,width=18cm,clip=}}
\caption[]{{\bf a.} Distribution of separations $d_{\rm min}$ of the 2540 runs for which both objects,  
\psr\ and HIP 99580 were situated between 25~pc and 35~pc from the centre of Cyg OB9. 
Drawn as solid line is a theoretical curve for a distribution of absolute differences between two Gaussian 
distributed 3D quantities with $\mu=0$ and 
$\sigma=5.2$~pc (see equation 2 in Tetzlaff et al. 2011). 
{\bf b.} Distribution of corresponding flight times $\tau_{kin}$ in the past since the SN.}
\label{histo}
\end{figure*}

\clearpage
\begin{figure*}
\centerline{\psfig{figure=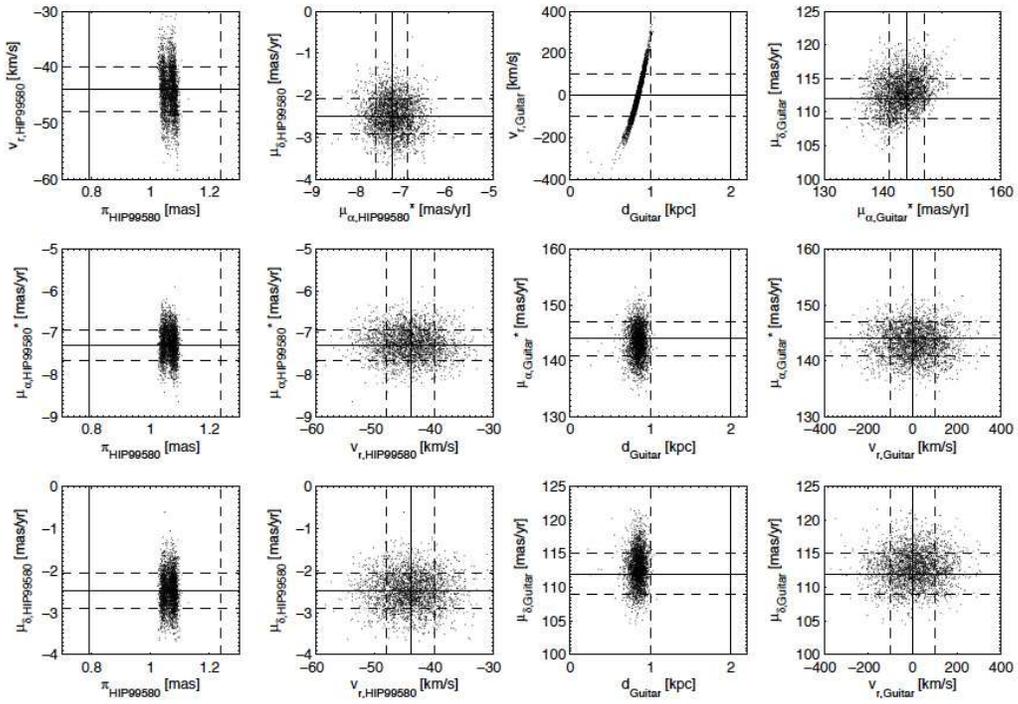,width=14cm,clip=}}
\caption[]{Input parameters (parallax $\pi$ or distance $d$, proper motion components 
$\mu_\alpha^\ast=\mu_\alpha\cos\delta$ and $\mu_\delta$, radial velocity $v_r$) 
for the Guitar pulsar and the runaway star HIP 99580 for the 2540 runs for which 
both objects were situated between 25~pc and 35~pc from the centre 
of Cyg OB9. The lines mark the currently accepted parameter values (solid) and their $1\sigma$ error bars (dashed).}
\label{correlation}
\end{figure*}

\clearpage
\begin{figure*}
\centerline{\psfig{figure=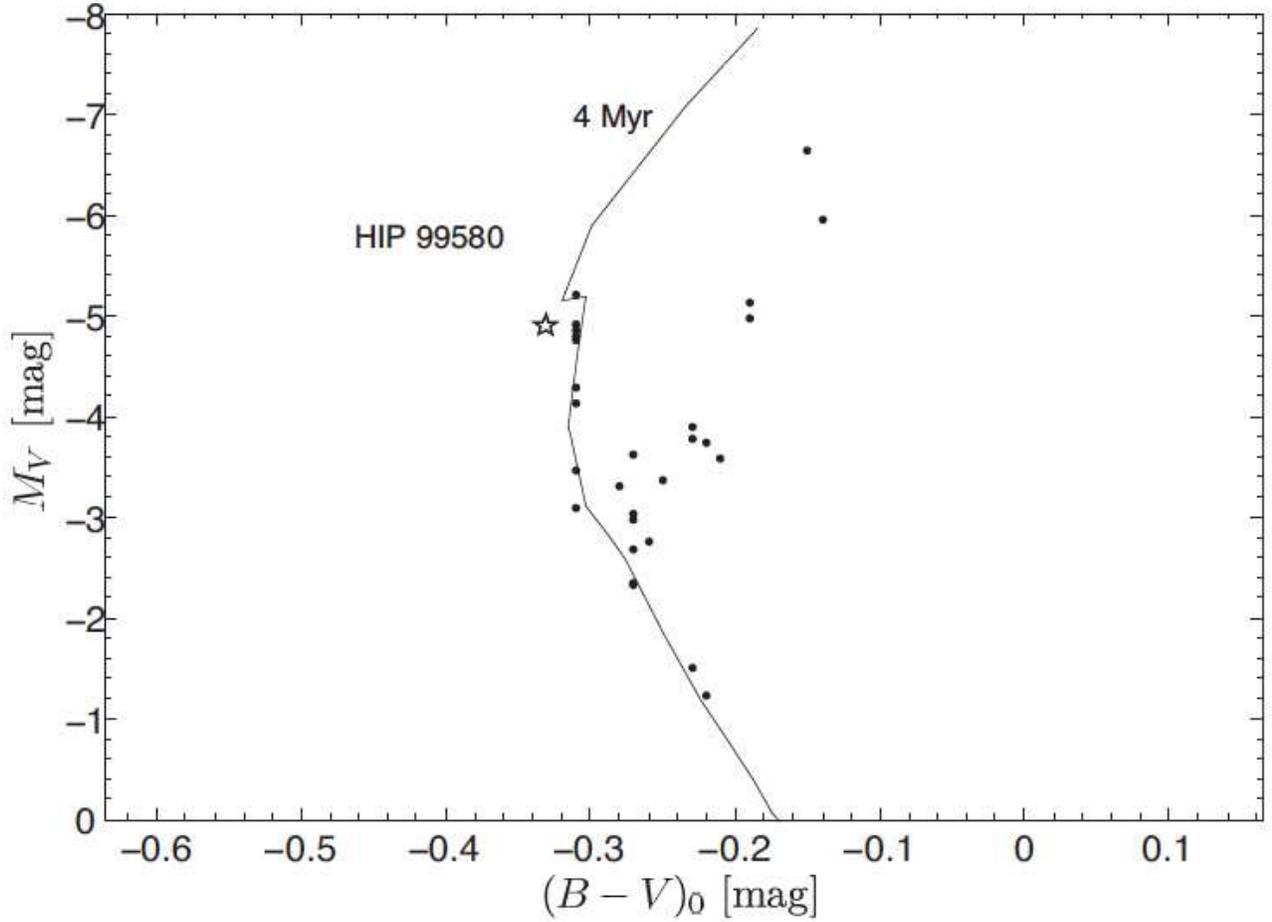,width=18cm,clip=}}
\caption[]{$\left(B-V\right)_0$ versus $M_V$ diagram of Cyg OB9 (dots represent member stars according to 
Garmany \& Stencel 1992), the potential parent association of the Guitar pulsar and the runaway star HIP 99850 (star). 
The solid line represents the 4~Myr isochrone from Marigo et al. (2008)  
(for solar metallicity; http://stev.oapd.inaf.it/cgi-bin/cmd). $\left(B-V\right)_0$ and $M_V$ of HIP 99580 are 
derived from its spectral type according to Schmidt-Kaler (1982).}
\label{bluestraggler}
\end{figure*}

\end{document}